\documentclass[12pt]{article} 
\usepackage{a4} 
\usepackage{color} 
\usepackage{epsfig}
\usepackage{amssymb} 
\usepackage{amsmath} 
\usepackage{amsfonts}

\newcommand{\news}{\setcounter{equation}{0}}

\newcommand{\X}{{\mathsf{H}}}

\newcommand{\R}{{\mathbb{R}}}

\newcommand{\C}{{\mathbb{C}}}
\newcommand{\CP}{{\mathbb{C}}{{P}}}

\newcommand{\beq}{\begin{equation}}
\newcommand{\eeq}{\end{equation}}
\newcommand{\bea}{\begin{eqnarray}}
\newcommand{\eea}{\end{eqnarray}}
\newcommand{\ben}{\begin{eqnarray*}}
\newcommand{\een}{\end{eqnarray*}}
\newcommand{\ra}{\rightarrow}

\newcommand{\cd}{\partial}

\newcommand{\less}{\backslash}
\newcommand{\M}{{\sf M}}

\def \d{\mathrm{d}}

\newcommand{\dstar}{\delta}
\newcommand{\ip}[1]{\langle #1 \rangle}

\newcommand{\oa}{\chi}
\newcommand{\kf}{\omega}

\newcommand{\ol}{\overline}

\begin{document}
\title{
  \vskip 15pt
  {\bf \large \bf Exact moduli space metrics for hyperbolic vortices}
  \vskip 10pt}
\author{
Steffen Krusch\thanks{E-mail: {\tt S.Krusch@kent.ac.uk}} \\[5pt]
{\normalsize {\sl Institute of Mathematics, Statistics and Actuarial 
Science}}\\
{\normalsize {\sl University of Kent,
Canterbury CT2 7NF, United Kingdom}} \\ \\
J.M. Speight\thanks{E-mail: {\tt speight@maths.leeds.ac.uk}}\\ [5pt]
{\normalsize {\sl Department of Pure Mathematics, University of Leeds}}\\
{\normalsize {\sl Leeds LS2 9JT, United Kingdom}}
}

\date{June 10, 2009}

\maketitle

\begin{abstract}
Exact metrics on some totally geodesic submanifolds of the
moduli space of static hyperbolic $N$-vortices are derived. These
submanifolds, denoted $\Sigma_{n,m}$, are spaces of $C_n$-invariant 
vortex configurations with $n$ single vortices at the vertices
of a regular polygon and $m=N-n$ coincident vortices at the polygon's centre.
The geometric properties of $\Sigma_{n,m}$ are investigated, and
it is found that $\Sigma_{n,n-1}$ is isometric to
the hyperbolic plane of curvature $-(3\pi n)^{-1}$. Geodesic flow on 
$\Sigma_{n,m}$, and a geometrically natural variant of geodesic flow
recently proposed by Collie and Tong, are analyzed in detail.

\end{abstract}

\vspace{2cm} 
\noindent
MSC classification numbers: 53C55; 53C80\newline


\newpage

\section{Introduction}
\news

Many aspects of the dynamics of topological solitons of Bogomol'nyi type
can be understood in terms of the geometry of the moduli space
$\M_N$ of static $N$-solitons. This approach, originally due to Manton,
addresses such diverse issues as low energy soliton scattering,
the thermodynamics of soliton gases, and the quantum mechanics of solitons.
For a comprehensive review, see \cite{Manton:2004tk}.
Mathematically, the main object of study is the $L^2$ metric $\gamma$,
a Riemannian metric on $\M_N$ which can be thought of
as the restriction of the kinetic energy of the parent field theory.
There are comparatively few situations in which explicit formulae for $\gamma$
are known, and one usually must make do with only partial or qualitative
information. 

This paper considers one of the rare cases where
explicit progress is possible, namely Ginzburg-Landau vortices 
moving on the hyperbolic plane. 
In this case, the Bogomol'nyi equations for static $N$-vortices can be
reduced to Liouville's equation on a disk, which is integrable.
Strachan exploited this fact \cite{Strachan:1992fb} to
obtain an implicit formula for $\gamma$ in terms of the analytic behaviour of
the Higgs field near the vortex centres. From this he deduced
explicit formulae for the metric on $\M_1$ and $\M_2$, but the calculations 
become intractable for $N\geq 3$. In this paper, we will find exact 
formulae for the induced metric on certain totally geodesic submanifolds
of $\M_N$ for all $N$, obtained by imposing invariance under certain
symmetry groups. Each submanifold $\Sigma_{n,m}\subset \M_N$ has (real)
dimension 2 and consists of static $N$-vortex solutions wherein
$n$ single vortices occupy the vertices of a regular polygon, and
$0\leq m\leq n-1$ coincident vortices sit at the polygon's centre (so $m+n=N$).
The two dimensions correspond to the orientation and radius of the polygon.
Geodesics in $\Sigma_{n,m}$ are conjectured to correspond to low-energy
$N$-vortex scattering trajectories in the case of slow, rotationally
equivariant initial data. 

We will discuss the curvature properties of $\Sigma_{n,m}$ and show that
$\Sigma_{n,n-1}$ is isometric to the hyperbolic
plane of curvature $-\frac{1}{3\pi n}$. This fact was already known 
(and is rather trivial) in the case $n=1$,
but its generalization to any $n\geq 1$ is new and extremely surprising.
It follows that
$\Sigma_{n,n-1}$ is isometric to a hyperboloid of
one sheet in $(2+1)$-dimensional Minkowski space
 (one of the standard models of hyperbolic space).
It turns out that, for $m$ sufficiently close to but different from
$n-1$, $\Sigma_{n,m}$ 
can still be isometrically embedded
as a surface of revolution in $\R^{(2,1)}$, and we construct the
generating curves for some of these surfaces numerically.
Geodesic motion on $\Sigma_{n,n-1}$ can be understood very directly, and
we obtain explicit formulae fixing the relationship between scattering angle
and impact parameter for two--vortex scattering with a stationary third vortex
at the origin (geodesic motion in $\Sigma_{2,1}$). Finally, we consider 
a variant of the moduli space dynamics recently proposed (for Euclidean
vortices) by Collie and Tong \cite{Collie:2008mx} in which the vortices
experience an effective magnetic field determined by the Ricci
curvature of $\M_{N}$. This flow is analyzed numerically on
$\Sigma_{2,0}$ and exactly on $\Sigma_{n,n-1}$.

\section{Vortices on the hyperbolic plane}
\news

In this section we review Ginzburg-Landau vortices on the hyperbolic plane of 
curvature $-\frac12$
\cite{Strachan:1992fb,Manton:2004tk}, which we denote $\X$, with metric $G$. 
It is convenient to use the Poincar\'e disk model of $\X$, so 
$\X=\{z\in\C\: :\: |z|<1\}$ and
\beq
G=\Omega\ dz d\ol{z},\qquad
\Omega=\frac{8}{(1-|z|^2)^2}.
\eeq
Then (critically coupled)
Ginzburg-Landau vortices on $\X$ are minimals of the potential
energy
\begin{equation}
\label{V}
V(\phi,A) = \frac{1}{2} \int_\X \left(
\d A\wedge\ast \d A +\overline{\d_A\phi}\wedge \ast \d_A\phi+
\frac{1}{4} (1- \bar{\phi} \phi)^2 \ast1\right), 
\end{equation}
where $\phi:\X\ra\C$ is a complex scalar field, $A\in\Omega^1(\X)$
is the gauge potential one-form, $\d_A\phi=\d\phi-iA\phi$, and $\ast$ is
the Hodge isomorphism.
The standard Bogomol'nyi argument shows that, among fields satisfying the
boundary condition $|\phi|=1$ for $|z|=1$, with winding number $N\geq 0$,
the potential energy $V$ satisfies
$$V\geq\pi N,$$
with equality if and only if
\begin{eqnarray}
\label{Bog1}
(\d_A\phi)\left(\cd/\cd\ol{z}\right) &=& 0, \\
\label{Bog2}
\d A - \ast\frac{1}{2} (1- \bar{\phi} \phi) &=& 0.
\end{eqnarray}
Hence solutions of (\ref{Bog1}), (\ref{Bog2}) minimize $V$ in their
homotopy class. Such solutions are called $N$-vortices, and the zeros
of $\phi$ are interpreted as individual vortex positions.

Equations (\ref{Bog1}) and (\ref{Bog2}) can be reduced to a single
gauge invariant equation by setting
$\phi = {\rm e}^{\frac{1}{2}h+i\chi}$. One obtains
\begin{equation}
4\frac{\cd^2h}{\cd z\cd\ol{z}} + \Omega - \Omega {\rm e}^h = 0.
\end{equation}
Setting $h = 2 g + 2\log \frac{1}{2}(1-|z|^2)$ the
equation for $h$ becomes Liouville's equation,
\begin{equation}
4\frac{\cd^2g}{\cd z\cd\ol{z}} - {\rm e}^{2g} = 0,
\end{equation}
which can be solved exactly. The solution is
\begin{equation}
g = - \log \frac{1}{2}\left(1-|f|^2\right) + \frac{1}{2} \log 
\left|\frac{df}{dz}\right|^2,
\end{equation}
where $f(z)$ is an arbitrary, complex analytic function. With a simple
choice of phase the scalar field is given by
\begin{equation}\label{phi}
\phi = \frac{1 - |z|^2}{1-|f|^2}\frac{df}{dz}.
\end{equation}
Then the  first Bogomol'nyi equation (\ref{Bog1}) is satisfied, if  and only if
\begin{equation}
A = -\ast\d \log \left(
\frac{1-|z|^2}{1-|f|^2} \right).
\end{equation}
Note that $\phi$ vanishes at the zeros of $\frac{df}{dz}$.

To ensure that $|\phi|\ra 1$ as $|z|\ra 1$, $\phi$ is nonsingular for $|z|<1$
and has winding number $N$, one must choose
\begin{equation}
\label{f(z)}
f(z) = \prod\limits_{i=1}^{N+1} \left( \frac{z - a_i}{1 - {\bar a}_i z}
\right),
\end{equation}
where $a_1,\ldots,a_{N+1}$ are arbitrary complex constants with $|a_i| < 1$.
Naively, it seems that the moduli space of static $N$-vortices should have
complex dimension $N+1$, but this overcounts, since meromorphic functions
$f(z)$ and gauge equivalence classes of solutions of the Bogomol'nyi
equations are not in one-to-one correspondence. In fact, the transformation
\beq
\label{symc}
 f \mapsto \frac{f-c}{1-{\bar c} f}, 
\eeq
where $|c|<1$ leaves $(\phi,A)$ unchanged up to gauge. One can use this 
freedom to set $a_{N+1}=0$, so
\begin{equation}
\label{f0}
f(z) = z \prod\limits_{i=1}^{N} \left( \frac{z - a_i}{1 - {\bar a}_i z}
\right),
\end{equation}
where $|a_i| <1$. Note that the denominator of this rational map is
uniquely determined by its numerator. So, hyperbolic $N$-vortices are
in one-to-one correspondence with degree $N$ polynomials
\beq
P(z)=\prod_{i=1}^N(z-a_i)
\eeq
all of whose roots lie in the open unit disk, and one concludes that
$\M_N$ has real dimension $2N$. The coefficients $a_i$ are
{\em not} natural complex coordinates on $\M_N$, however, as we shall see.

It is useful to think of $f$ as a holomorphic mapping of the Riemann
sphere $\C\cup\{\infty\}$ to itself. Clearly $f$ has exactly $2N$ critical 
points, counted with multiplicity. Note that $f$ commutes with the 
involution $\iota:z\mapsto\bar{z}^{-1}$ (reflexion in the equator) so if
$z$ is a critical point of $f$, so is $\iota(z)$. Since $f$ has no critical 
points on the equator, $|z|=1$, it follows that it has exactly $N$
critical points inside the unit disk, and $N$ outside. Let us denote the
$N$ (not necessarily distinct) critical points inside the unit disk
$Z_r$, $r=1,2,\ldots,N$. These are interpreted as the vortex positions.
They provide local complex coordinates on $\M_N\less\Delta_N$, where
$\Delta_N$ is the coincidence set, that is, the set of $N$-vortices for which
the vortex positions are not all distinct. Good {\em global} complex
coordinates 
are provided by the coefficients of the polynomial $\prod_{r=1}^N(z-Z_r)$.
This equips $\M_N$ with a canonical complex structure which coincides, off
$\Delta_N$, with the structure defined by the coordinates $Z_r$. Note that
the critical points $Z_r$ of $f$ depend non-holomorphically on the
parameters $a_i$, so this canonical complex structure is different from
the complex structure defined by the coordinates $a_i$ on $\M_N$. 

There is a natural Riemannian metric $\gamma$ on $\M_N$ defined by restricting 
the kinetic energy of the abelian Higgs model
\beq
T=\frac12\int_\X\left(\cd_0 A\wedge\ast\cd_0A+|\cd_0\phi|^2\ast1\right)
\eeq
to $T\M_N$. We here choose to work in temporal gauge ($A_0=0$), so must
impose Gauss's law 
\beq
\dstar\cd_0 A=\frac{i}{2}(\ol{\phi}\cd_0\phi-\phi\cd_0\ol{\phi})
\eeq
as a constraint on $(\cd_0\phi,\cd_0A)$, where $\dstar=-\ast\d\ast$ is the
coderivative on $(\X,G)$. It is known that $\gamma$ is K\"ahler with
respect to  
the canonical complex structure just defined. This follows from a formula
for $\gamma$ first obtained by Strachan \cite{Strachan:1992fb} and
later generalized and reinterpreted by Samols \cite{Samols:1991ne} and 
Romao \cite{romthe}.
The formula gives $\gamma$ on $\M_N\less\Delta_N$. In the next section, we
will
need to compute the induced metric on certain totally geodesic submanifolds
$\Sigma_{n,m}\subset\M_{n+m}$ which (for $2\leq m\leq n-1$) lie
entirely in $\Delta_{n+m}$, so we must make a small modification to 
Samols' calculation.

Rather than compute $\gamma$ on $\M_N\less\Delta_N$, we fix the
positions of $m$ of the vortices (not necessarily distinct) and allow
only the other $n=N-m$ vortices to move, along trajectories
$Z_1(t),\ldots,Z_r(t)$ which remain distinct from one another, and the fixed 
zeros. This, then, defines an $n$-dimensional complex submanifold of
$\M_{n+m}$, which we denote by $\M_n^{p(z)}$, where $p(z)$ is the unique
monic degree $m$ polynomial whose roots are the fixed zeros (so
$\M_n^1\equiv\M_n$). Samols' argument generalizes immediately to this setting,
and one finds that the kinetic energy of a trajectory in $\M_n^{p(z)}$ is
\begin{equation}
\label{kinetic}
T = \frac12\pi \sum\limits_{r,s=1}^n
\left( \Omega(Z_r) \delta_{rs} + 2 \frac{\partial b_s}{\partial Z_r}
\right) \dot{Z}_r \dot{{\bar Z}}_s.
\end{equation}
Here $b_r$ are coefficients in the
expansion of $h=\log|\phi|^2$ around the zero $Z_r$ of $\phi$,
\begin{equation}
\label{hb}
h = 2\log|z-Z_r|+a_r+\frac{1}{2}{\bar b}_r (z-Z_r)
+\frac{1}{2} b_r ({\bar z} - {\bar Z}_r) + \dots
\end{equation}
On hyperbolic space, $|\phi|^2$ is known explicitly, so the dependence of
$b_r$ on the vortex positions $Z_1,\ldots,Z_n$ can also be determined
explicitly, in principle. In practice, as we shall see, this is
very difficult, but some properties of the coefficients $b_r$ are
immediate. First, since $T$ is manifestly real, one must have
\beq
\label{julbin}
\frac{\cd b_r}{\cd Z_s}\equiv\frac{\cd\bar{b}_s}{\cd\bar{Z}_r}.
\eeq
It follows that the metric on $\M_n^{p(z)}$ induced by $T$,
\beq
\label{metric}
\gamma = \pi\sum\limits_{r,s=1}^n
\left( \Omega(Z_r) \delta_{rs} + 2 \frac{\partial b_s}{\partial Z_r}
\right) dZ_r d{\bar Z}_s,
\eeq
is Hermitian with respect to the canonical complex structure, with
K\"ahler form
\beq
\label{kf}
\omega = \frac{i}{2}\pi\sum\limits_{r,s=1}^n
\left( \Omega(Z_r) \delta_{rs} + 2 \frac{\partial b_s}{\partial Z_r}
\right) dZ_r \wedge d{\bar Z}_s.
\eeq
Clearly, $\d\omega=0$, by (\ref{julbin}), that is, $\gamma$ is K\"ahler.
Following Romao, it is convenient to define the $(0,1)$ form 
\beq
\label{bformdef}
b=\sum_{r=1}^nb_rd\bar{Z}_r
\eeq
on $\M_n^{p(z)}$, which is known to satisfy $\bar\cd b=0,$ \cite{romthe}. The
K\"ahler form on $\M_n^{p(z)}$ may then be compactly written
\beq
\label{kfnice}
\omega=\frac{i}{2}\pi\left(2\d b+\sum_{r=1}^n\Omega(Z_r)dZ_r\wedge d\bar{Z}_r
\right).
\eeq

\section{Totally geodesic submanifolds of $\M_N$}
\news

We shall use formula (\ref{kfnice}) to deduce the induced metric on certain
totally geodesic submanifolds of $\M_N$ which we now define.
Recall that points in $\M_N$ are in one-to-one correspondence with
monic degree $N$ polynomials with roots only in the unit disk.
We can choose to identify a $N$-vortex with the polynomial whose roots are the
vortex positions, but it is somewhat more convenient to identify it 
instead with the monic polynomial $P(z)$ defined by the numerator
of the rational map $f(z)$, whose coefficients we denote $\beta_i\in\C$,
\beq
P(z)=z^N+\beta_1z^{N-1}+\cdots+\beta_N.
\eeq
There is an obvious action of $U(1)$ on such polynomials, whose associated
action on $\M_N$ is clearly isometric (it just rotates the vortex positions
about $z=0$), namely,
\beq
\label{U(1)}
P(z)\mapsto \lambda^{-N}P(\lambda z)
\eeq
for each $\lambda\in U(1)$. In terms of the coefficients of $P(z)$, the action
is
\beq
(\beta_1,\beta_2,\ldots,\beta_N)\mapsto 
(\lambda^{-1}\beta_1,\lambda^{-2}\beta_2,\ldots,\lambda^{-N}\beta_N).
\eeq
From now on we consider the case where $\lambda=e^{2\pi i/n}$, for $n$
some positive integer, 
so that $\lambda$ generates the cyclic group $C_n$. Then $P(z)$ is a fixed
point of $\lambda$ if and only if $\beta_k=0$ for all $k$ not divisible by
$n$. So the fixed point set of $\lambda$ consists of all
polynomials of the form
\beq
P(z)=z^N+\beta_nz^{N-n}+\beta_{2n}z^{N-2n}+\cdots+\beta_{pn}z^{N-pn},
\eeq
where $p=\left\lceil\frac{N}{n}\right\rceil$. 
The corresponding submanifold of $\M_N$ has complex dimension $p$ and 
is totally geodesic by a standard symmetry argument \cite{Romao:2004gc}. 
In particular, if $\frac{N}{2}<n\leq N$, so that
$p=1$, the fixed point set has complex dimension one and consists of
polynomials
\beq
\label{F(z)}
P(z)=z^m(z^n+\beta_n),\qquad |\beta_n|<1, \qquad 0 \le m < n.
\eeq
Let us denote this totally geodesic submanifold $\Sigma_{n,m}\subset\M_{n+m}$.
In order to calculate the metric on $\Sigma_{n,m}$ we first need to find the
vortex positions, that is, the 
critical points of 
\beq
f(z) = \frac{z^{m+1} \left(z^n - a^n\right)}{1 - {\overline a}^n z^n}, 
\eeq
which arises from \eqref{F(z)} (replacing $\beta_n = - a^n$ for later
convenience). Recall $f$ has exactly $n+m$ critical points, counted
with multiplicity, in the unit disk $\X$. Now $f(\lambda z)\equiv\lambda^{m+1}
f(z)$, so $z$ is a critical point of $f$ if and only if $\lambda z$ is a
critical point. Hence, $f$ has $n$ critical points $Z_r$ at the
vertices of some regular $n$-gon,
\begin{equation}
\label{zeros}
Z_r = \alpha~\lambda^{r-1},
\end{equation}
and the other $m$ critical points must be coincident at $0$ (the only fixed
point in $\X$ of $\lambda$). So $\Sigma_{n,m}$ consists of vortex
configurations with $m$ vortices coincident at $z=0$
and $n$ vortices located at the vertices of a regular polygon centred
on $0$. 

We next determine how the vortex position $\alpha=Z_1\in\C$
is related to the complex parameter $a$. If $a$ is real positive, then
$f$ clearly has a real postive critical point, and $f$ remains unchanged under
$a\mapsto\lambda a$. It follows that 
\beq
a = \alpha \nu(|\alpha|)
\eeq
where $\nu$ is a positive real function of $|\alpha|$ only.
 A short calculation leads to a
quadratic equation in $\nu^n$. We choose the solution which
satisfies $|a| < 1$ provided $|\alpha|<1$ and obtain
\begin{equation}
\label{lambda}
\nu^n = \frac{(m+1) \left(1+ |\alpha|^{2n} \right)
-\sqrt{(m+1)^2 \left(1-|\alpha|^{2n} \right)^2 + 4 n^2 |\alpha|^{2n}}}
{2 |\alpha|^{2n} \left(m+1-n \right)}
\end{equation}
for $n > m+1$. For $n = m+1$ the solution simplifies considerably, namely, 
\begin{equation}
\label{lambda1}
\nu^n = \frac{2}{1+|\alpha|^{2n}}.
\end{equation}  

We can now eliminate the parameter $a$ from $f(z)$ in favour of $\alpha$ which,
being one of the vortex positions, is a good (local) complex
coordinate on $\Sigma_{n,m}$ with respect to the canonical complex structure:
\beq
f(z) = \frac{z^{m+1} (z^n - \nu^n \alpha^n)}{1-\nu^n {\bar
    \alpha}^n z^n}.
\eeq
By the K\"ahler property and rotational invariance, we know
 that the metric on $\Sigma_{n,m}$ must take the form
\beq
\gamma=F(|\alpha|)d\alpha d{\bar \alpha}
\eeq
for some conformal factor $F(|\alpha|)$.
To compute $F$, we require information on the coefficients
$b_r$, and hence a formula for $h=\log|\phi|^2$. Now $f'(z)$ has
zeros at $\lambda^r \alpha$ and $\lambda^r\bar{\alpha}^{-1}$ (recall critical
points of $f$ are paired $z\leftrightarrow \bar{z}^{-1}$), and a zero of order
$m$ at $0$. It follows that
\beq
f'(z) = \frac{(m+1) \nu^n z^m(z^n-\alpha^n)(1-{\bar
    \alpha}^n z^n)}{(1- \nu^n {\bar \alpha}^n z^n)^2}.
\eeq
In order to avoid the logarithmic singularities of $h$ near
$z=Z_1=\alpha$, we define the regularized version of \eqref{hb} as  
\beq
h_{reg}
= \log |\phi|^2 -\log \left(z - \alpha\right) 
- \log \left({\bar z} - {\bar \alpha}\right). 
\eeq
Since $\alpha$ is a simple zero we can calculate the coefficient $b_1$
in \eqref{hb} as
\beq
\label{bb}
b_1 = \left. \frac{2 \partial h_{reg}}{\partial {\bar z}}\right|_{z=\alpha},
\eeq
which leads to
\beq
\label{b}
b_1 = \frac{1}{{\bar \alpha}} \left(
2m+n-1 - \frac{2n\alpha^n {\bar \alpha}^n}{1-\alpha^n{\bar \alpha}^n}
+ \frac{4n \nu^n \alpha^n {\bar \alpha}^n}{1-\nu^n \alpha^n
  {\bar \alpha}^n} - \frac{4 \alpha {\bar \alpha}}{1- \alpha {\bar
    \alpha}}\right).
\eeq

We have calculated $b_1$ only, but we can deduce $b_r$ for $r\geq 2$
by rotational symmetry.
From the definition of the coefficient $b_r$ in \eqref{hb}, it follows that
if we simultaneously rotate all the zeros by $\lambda'\in U(1)$, the
$b_r$ coefficients transform as
\beq\label{sym}
b_r(\lambda'Z_1,\ldots,\lambda'Z_n)=\lambda'b_r(Z_1,\ldots,Z_n).
\eeq
In the case $\lambda'=\lambda=e^{2\pi i/n}$, this rotation just cyclically
permutes the vortices, so we see that
\beq\label{mirric}
b_r(\alpha)=\lambda^{r-1}b_1(\alpha).
\eeq
It also follows from \eqref{sym} that $\bar{\alpha}b_1(\alpha)$ is a
function of $|\alpha|$ only, which is a consistency check on our formula
\eqref{b}.

To complete the calculation of the metric, we note that $\Sigma_{n,m}$
(with the point $\alpha=0$ removed, strictly speaking) is a submanifold
of $\M_n^{z^m}$, on which the formula \eqref{kfnice} for the K\"ahler
form holds. Let us introduce the real vector fields\footnote{In
  \cite{Chen:2004xu}, Chen and Manton used the vector field $X$ to
  derive an interesting integral formula for the K\"ahler potential.} 
\beq
X=\sum_{r=1}^n\left(Z_r\frac{\cd\:}{\cd
  Z_r}+\bar{Z}_r\frac{\cd\:}{\cd\bar{Z}_r} \right),\qquad
Y=JX=i\sum_{r=1}^n\left(Z_r\frac{\cd\:}{\cd Z_r}-
\bar{Z}_r\frac{\cd\:}{\cd\bar{Z}_r}\right)
\eeq
on $\M^{z^m}_n$, and note that on $\Sigma_{n,m}$ these coincide with
\beq
X=|\alpha|\frac{\cd\:\:}{\cd|\alpha|},\qquad
Y=\frac{\cd\:}{\cd\psi},
\eeq
where $\alpha=|\alpha|e^{i\psi}$. Hence, the conformal factor we seek is
\bea
F(|\alpha|)&=&\frac{1}{|\alpha|^2}\omega(X,Y)
=\frac{\pi}{|\alpha|^2}\Big(i(\d b)(X,Y)+n|\alpha|^2\Omega(|\alpha|)\Big)
\nonumber \\
&=&\frac{i\pi}{|\alpha|^2}\Big(X[b(Y)]-Y[(b(X)]\Big)+n\pi\Omega(|\alpha|)
\eea
since $[X,Y]=0$.
Now 
\beq
b(X)=\sum_{r=1}^nb_r d\bar{Z}_r(X)=\sum_{r=1}^n\bar{Z}_rb_r
\eeq
and $b(Y)=b(JX)=-ib(X)$ since $b$ is a $(0,1)$ form. Since $X$ and
$Y$ are tangential to $\Sigma_{n,m}$, to compute $X[b(Y)]=-iX[b(X)]$ and 
$Y[b(X)]$
it suffices to know $b(X)$ only on $\Sigma_{n,m}$, where $Z_r=\lambda^{r-1}
\alpha$. But using \eqref{mirric} we see that, on $\Sigma_{m,n}$,
$b(X)=n\bar{\alpha}b_1(\alpha)$ which, as we have remarked, is a function
of $|\alpha|$ only. Hence, $Y[b(X)]=0$, and we find that
\beq\label{met}
F(|\alpha|)=\pi n\left(\Omega(|\alpha|)+\frac{1}{|\alpha|}
\frac{d\:\:\:}{d|\alpha|}(\bar{\alpha}b_1(\alpha))\right).
\eeq
The derivation for this formula used only rotational invariance and
\eqref{kfnice}, so it holds equally well for vortex polygons on the
Euclidean plane, or any other surface of revolution.

Now, we can evaluate the metric 
using equations \eqref{met},
\eqref{b}, \eqref{lambda} and \eqref{lambda1},
obtaining
\beq
\label{m1}
\gamma = 
\frac{4 \pi n^3 |\alpha|^{2n-2}   }{(1-|\alpha|^{2n})^2}
\left(1 + \frac{2n(1+|\alpha|^{2n})}{\sqrt{(m+1)^2(1-|\alpha|^{2n})^2 +
    4 n^2 |\alpha|^{2n}}}
\right) d\alpha d{\bar \alpha}
\eeq 
for $n>m+1.$ For $n=m+1$, the metric simplifies and we have
\beq
\label{m2}
\gamma = \frac{12 \pi n^3 |\alpha|^{2n-2}}{(1-|\alpha|^{2n})^2}\,
    d\alpha d{\bar \alpha}. 
\eeq
In the case $m=0$, equation \eqref{m1} agrees with previous results of
Strachan \cite{Strachan:1992fb} for $n=1,2$, and Krusch and Sutcliffe
\cite{Krusch:2005wr} for general $n$. For $|\alpha|$ close to $1$, the 
moving vortices are very far apart, so one expects the metric to approach the
metric induced on $\Sigma_{m,n}$ by the product metric on $(\M_1)^n$
\beq
\gamma_{\rm product}=\sum_{r=1}^n\frac{12\pi}{(1-|Z_r|^2)^2} dZ_rd\bar{Z}_r,
\eeq
which is 
\beq
\gamma_\infty=\frac{12\pi n}{(1-|\alpha|^2)^2} d\alpha d\bar{\alpha},
\eeq
the metric on the hyperbolic plane of curvature $\kappa=-\frac{1}{3\pi n}$.
Indeed, the full metric {\em is} asymptotic to $\gamma_\infty$:
\beq
\gamma=\left(\frac{12\pi n}{(1-|\alpha|^2)^2}
-nm(m+2)\pi+O(1-|\alpha|)\right)d\alpha d\bar\alpha.
\eeq
Much more surprising is the fact that $\Sigma_{n,n-1}$ is {\em exactly}
 isometric
to the hyperbolic plane of curvature $\kappa=-\frac{1}{3\pi n}$.
To see this, one should introduce the global complex coordinate
$\oa=\alpha^n$, which is in one-to-one correspondence with points in
$\Sigma_{n,m}$ (recall that $\alpha\mapsto\lambda\alpha$ just permutes the
vortex positions, so maps each point in $\Sigma_{n,m}$ to itself).
The metric is then 
\beq\label{alberta}
\gamma =  
\frac{4 \pi n}
{\left(1-|\oa|^{2} \right)^2}
\left( 1 +
\frac{2n \left(1 + |\oa|^{2} \right)}
{\sqrt{(m+1)^2\left(1-|\oa|^{2} \right)^2 + 4 n^2 |\oa|^{2}}}
\right)
d\oa d{\bar \oa}.
\eeq
Note that this is nondegenerate at $\oa=0$.
For $m=n-1$, it simplifies to
\begin{equation}
\gamma = \frac{12 \pi n \ d\oa d{\bar \oa}}
{\left(1-|\oa|^{2} \right)^2},
\end{equation}
which, as promised, is the metric on the  hyperbolic plane with
curvature $\kappa = -\frac{1}{3 \pi n}$.

\section{The geometry of $\Sigma_{n,m}$}
\news

It follows immediately from (\ref{alberta}) that $\Sigma_{n,m}$ is 
geodesically complete and has infinite volume. 
The radial curves
$\oa(t)=t\lambda$, $\lambda\in U(1)$, $-1<t<1$ are pregeodesics corresponding
to the dual polygon scattering trajectories characteristic of 
topological solitons in two dimensions. It is straightforward to compute
the Gauss curvature of the surfaces $\Sigma_{m,n}$ using the formula
\beq
\kappa\left(|\oa|\right)=-\frac{1}{2|\oa|F(|\oa|)}\frac{d\,\,}{d|\oa|}
\left(\frac{|\oa|}{F(|\oa|)}\frac{dF}{d|\oa|}\right).
\eeq
In all cases $$\lim\limits_{|\oa|\ra 1}\kappa=-\frac{1}{3 \pi n},$$ and
for $m$ close to $n-1$, $\kappa$ is uniformly negative. As already
noted, for $m=n-1$, 
$\kappa=-\frac{1}{3 \pi n}$ for all $\oa$.
By contrast, for $m$ small, 
$\kappa$ is positive close to $\oa=0$. For example, on
$\Sigma_{n,0}$ one has
\beq
\kappa(0)={\frac {2\,{n}^{3}-4\,n-1}{\pi \,n \left( 1+4\,n+4\,{n}^{2}
    \right) }} 
\eeq
which is positive for all $n\geq 2$. 

\begin{figure}
\begin{center}
\includegraphics[scale=0.5]{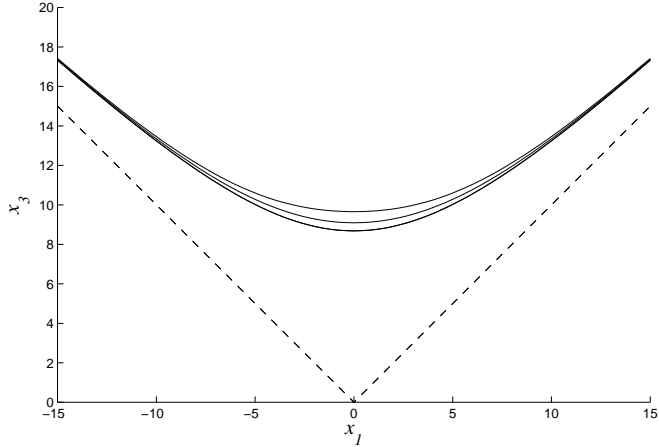}
\end{center}
\caption{
Generating curves for the
 spaces $\Sigma_{8,m}$ isometrically embedded as surfaces
of revolution in $\R^{(2,1)}$ for $m=5,6,7$ (solid curves, top to bottom).
The full surfaces are swept out by rotating about the $x_3$ axis.
The $m=7$ curve coincides with the hyperbola $x_1^2-x_3^2=-24\pi $. The
dashed lines mark the light cone. No isometric embeddings of $\Sigma_{8,m}$
for  $m=0,\ldots,4$ exist since these have
positive Gauss curvature at the fixed point of the rotation action.}
\label{fig1}
\end{figure}

For purposes of visualization, it would be useful to find an isometric
embedding of $\Sigma_{m,n}$ as a surface of revolution in $\R^3$, along the
lines of the ``rounded cone'' picture of the Euclidean two-vortex moduli space
discussed by Samols \cite{Samols:1991ne}. In our case, each $\Sigma_{n,m}$ is
asymptotic to a complete 
space of constant negative curvature, so it is clear that
no such isometric embedding exists. However, we
certainly can find an isometric embedding of $\Sigma_{n,n-1}$ as a surface of
revolution in $\R^{(2,1)}$ (that is, $\R^3$ equipped with the 
Lorentzian metric
$dx_1^2+dx_2^2-dx_3^2$), invariant under rotations about the
timelike $x_3$ axis. The image of this embedding is the hyperboloid
on which
\beq
x_1^2+x_2^2-x_3^2=-3\pi n,\qquad x_3>0.
\eeq
Of course, this is just the hyperboloid model of the hyperbolic plane.
One expects to be able to generalize this to $\Sigma_{n,m}$ at least for
$m$ close to $n-1$. A short calculation shows that any surface of
revolution which intersects the symmetry axis must have $\kappa\leq 0$ at
the intersection point, so isometric embeddings of $\Sigma_{n,m}$ for
$m$ small certainly do not exist, as $\kappa(0)>0$ and $\oa=0$ is
a fixed point of the rotation action. Using a straightforward modification
of the method described in \cite{McGlade:2005ug}, one can obtain
generating curves for the embedded surfaces of revolution for $m$ close to
$n-1$. Some examples are depicted in figure \ref{fig1}.

\begin{figure}
\begin{center}
\includegraphics[scale=0.5]{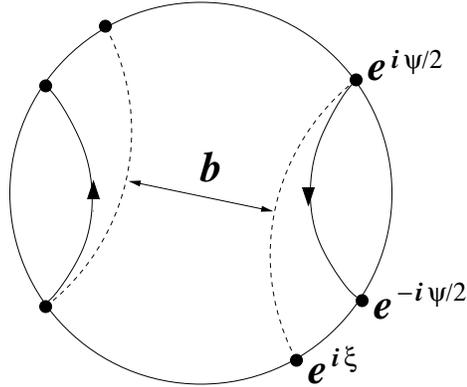}
\end{center}
\caption{The geometry of vortex scattering: the solid curves are the 
vortex trajectories in physical space, while the dashed curves are
the free trajectories which the vortices would describe if they did not
interact with one another. By definition, these are the geodesic
arcs which have second order contact with the incoming trajectories at
their entry points on the boundary of $\X$.}
\label{fig2}
\end{figure}

The geodesic flow on $\Sigma_{2,0}$ was considered in detail in 
\cite{Strachan:1992fb}. It is a simple matter to analyze geodesic
motion in $\Sigma_{n,n-1}$ for any $n\geq 1$, since each of these spaces is
isometric to a hyperbolic plane, and geodesic flow is invariant under 
homothety. Hence, the geodesic trajectories in $\Sigma_{n,n-1}$ are
the standard geodesics in the hyperbolic plane. In the Poincar\'e disk
model, these are circular (or straight) arcs which intersect the
boundary of the unit disk orthogonally. The vortex
trajectories in physical space $\X$ corresponding to a geodesic in 
$\Sigma_{n,n-1}$ are the preimage of the geodesic
under the map $z\mapsto z^n$ of the unit
disk to itself. Since this map is conformal, the vortex trajectories
also intersect the unit circle orthogonally. 
Consider the case $n=2$ in detail. Without loss of generality, we
may restrict attention to the geodesics in $\Sigma_{2,1}$ which intersect
the boundary at $e^{\pm i\psi}$ where $0<\psi\leq\frac\pi2$. Each geodesic
$\oa(t)$
corresponds to a 3-vortex motion in which one vortex remains stationary
at the origin and the other two move towards each other, scatter and recede,
following the trajectories
$z(t)=\pm\oa(t)^{\frac12}$.
We may derive an explicit formula for the scattering angle 
$\Theta$ and impact parameter $b$ associated with the $\psi$ geodesic
as follows. First we determine the geodesic in $\X$ which makes second
order contact with the incoming vortex trajectory where it intersects the
boundary, at $e^{i\psi/2}$. Simple trigonometry shows that this is a circular
arc of radius $\frac12\tan\psi$. 
This is the path that this vortex would follow if
it did not interact with the other two. Call its exit point
$e^{i\xi}$. Then the scattering angle is the angle one must rotate the
disk so that this free exit point shifts to the actual vortex exit point
$e^{-i\psi/2}$, that is, $\Theta$ such that $\xi+\Theta=-\frac\psi2$.
One can similarly construct the
initial free path 
of the other moving vortex: it is the image of the first under a rotation
by $\pi$
about the origin.
We define the impact parameter to be the 
hyperbolic distance
(with respect to metric $G$ on  $\X$) between these two free trajectories.
The geometry is summarized in figure \ref{fig2}. Straightforward
calculation shows that
\bea
b(\psi)&=&4\sqrt{2}\tanh^{-1}\left(\frac{\sqrt{4-3\sin^2\psi}-
  \sin\psi}{2\cos\psi}\right),\nonumber\\  
\Theta(\psi)&=&2\tan^{-1}\left(\frac12\tan\psi\right)-\psi.
\eea
Note that  
$\psi=\frac\pi2$ gives the expected $90^\circ$ head-on scattering process
(i.e.\ $b=0$ and $\Theta=\frac\pi2$), and that $\Theta$ decreases
monotonically towards $0$ as $b$ increases.

\section{The Collie-Tong flow}
\news

Motivated by a supersymmetric extension of the Abelian Higgs model with 
a Chern-Simons term, Collie and Tong have derived a new geometrically
natural moduli space dynamics for Euclidean vortices
\cite{Collie:2008mx}. This
geometric flow is well defined on any K\"ahler soliton moduli space, and
it is interesting to consider the dynamics of hyperbolic vortices 
under this flow, in comparison with their Euclidean counterparts. Recall
that on a K\"ahler manifold $\M$ one has a closed 2-form $\rho$ constructed
from the Ricci curvature $R$ and almost complex structure
$J$ in the same way that the K\"ahler form is
constructed from the metric and $J$, that is,
\beq
\rho(X,Y)=R(JX,Y).
\eeq
A trajectory $\oa(t)\in \M$ is a solution of the flow if
\beq\label{ctf}
\nabla_{d/dt}\dot{\oa}-\lambda\sharp\iota_{\dot{\oa}}\rho=0
\eeq
where $\lambda$ is a real constant, $\sharp$ denotes the metric isomorphism
$T^*\M\ra T\M$ and $\iota$ denotes interior product, 
$(\iota_X\rho)(Y)=\rho(X,Y)$. A very similar flow has, in fact, been 
derived previously by Kim and Lee \cite{Kim:2002qma}, but it is the
geometrically natural  
formulation given above which is significant for our purposes, and this
is due to Collie and Tong. For this reason, we will call equation (\ref{ctf})
the Collie-Tong flow. 
It follows from (\ref{ctf}) that
\beq
\frac{d}{dt}\ip{\dot\oa,\dot\oa}=
2\lambda\ip{\dot\oa,\sharp\iota_{\dot{\oa}}\rho}
=2\lambda\rho(\dot\oa,\dot\oa)=0,
\eeq
so this flow, like geodesic flow, conserves speed $\|\dot\oa\|$.
Given a solution $\oa(t)$ of (\ref{ctf}), $\oa(\lambda_0 t)$ satisfies
(\ref{ctf}) with parameter $\lambda'=\lambda/\lambda_0$, 
so we can rescale $\lambda$ to any convenient value. Note that, in contrast
to geodesic motion, the trajectories depend on $\|\dot\oa(0)\|$, not
just the direction of $\dot{\oa}(0)$. If $\lambda=0$, one recovers geodesic
flow, so one expects trajectories under (\ref{ctf}) to approach geodesics
when  the initial speed is taken to infinity.

\begin{figure}
\begin{center}
\begin{tabular}{cccc}
(a)&(b)&(c)&(d)\\
& & & \\
\includegraphics[scale=0.25]{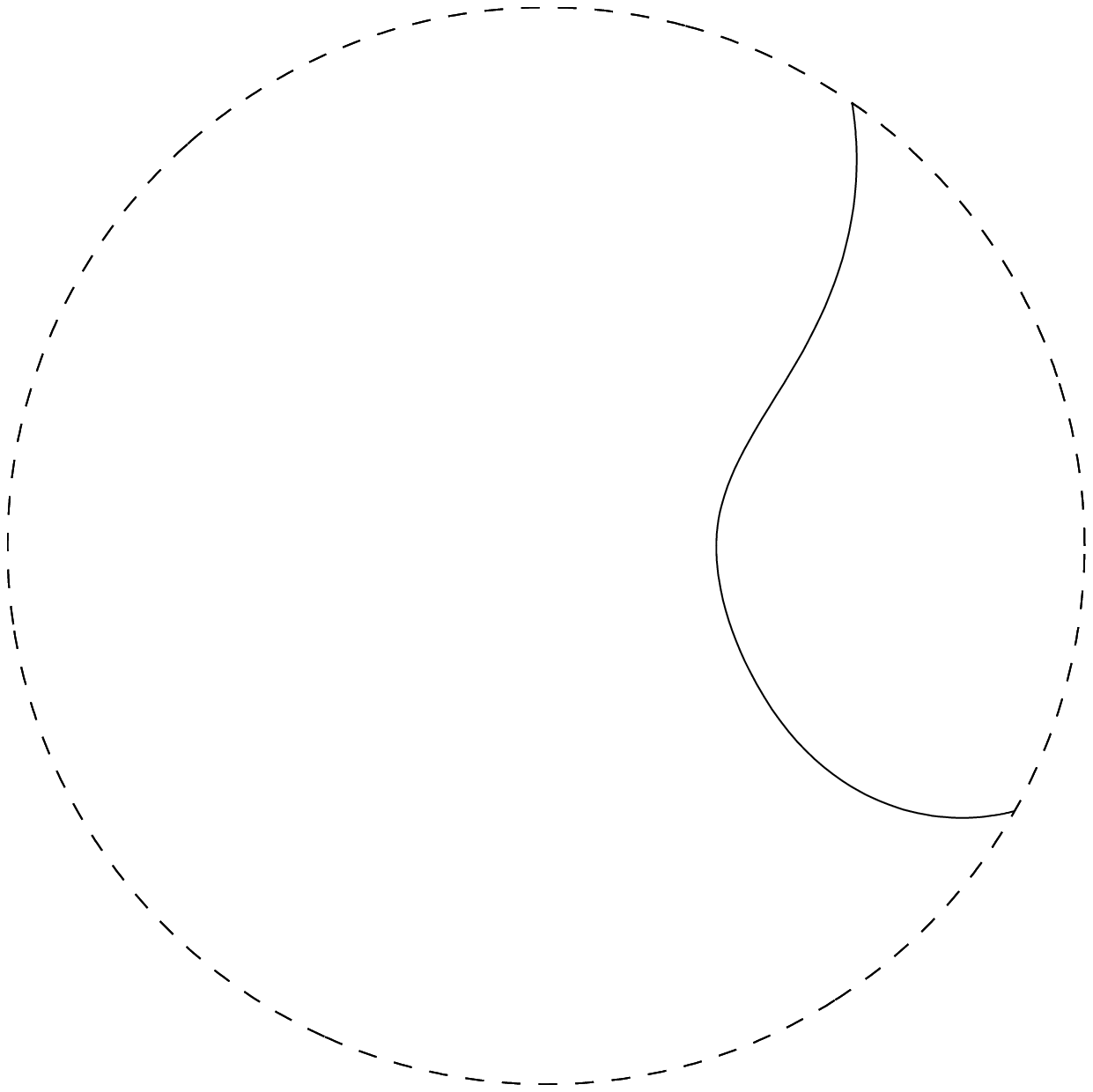}&
\includegraphics[scale=0.25]{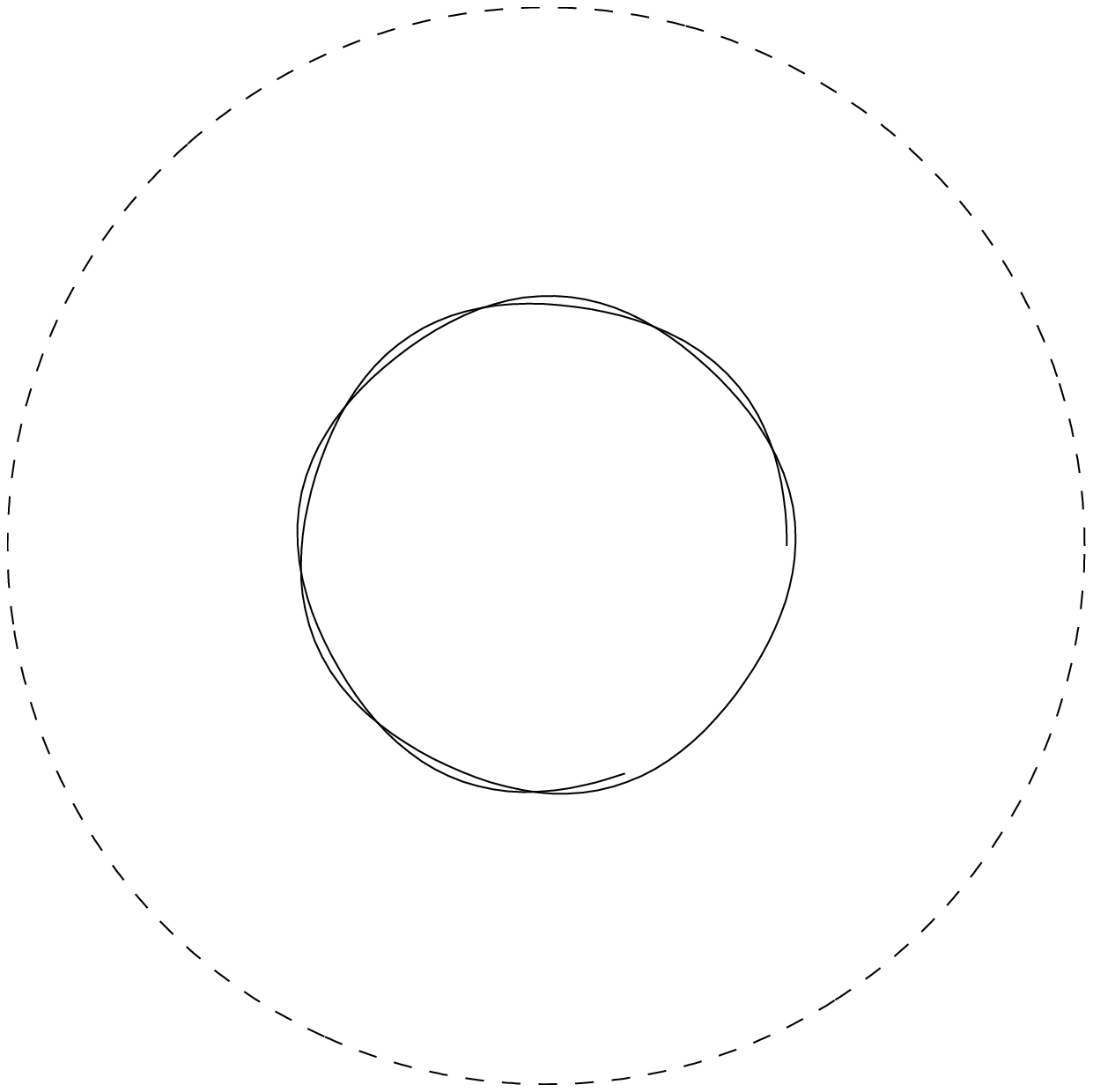}&
\includegraphics[scale=0.25]{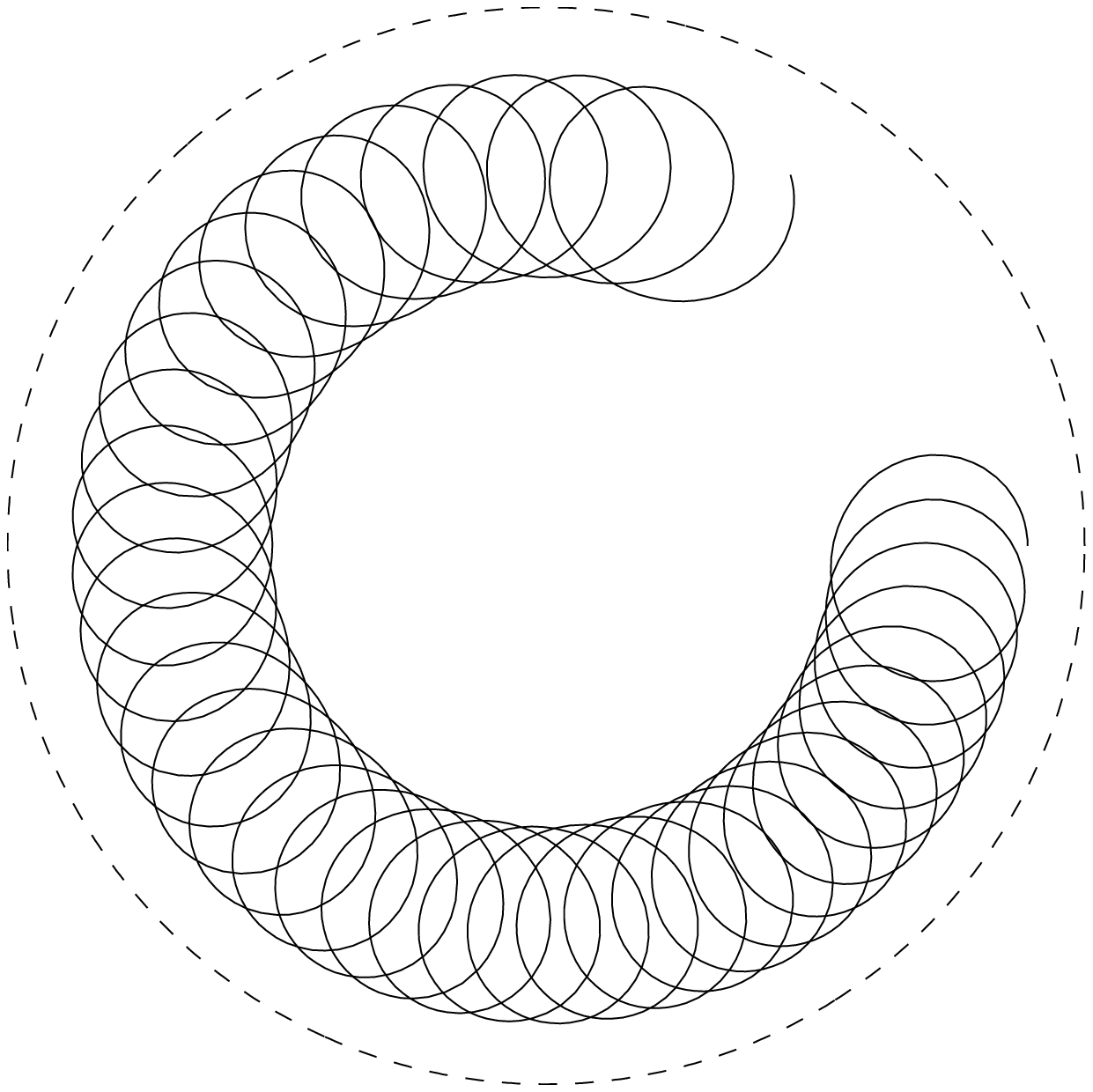}&
\includegraphics[scale=0.25]{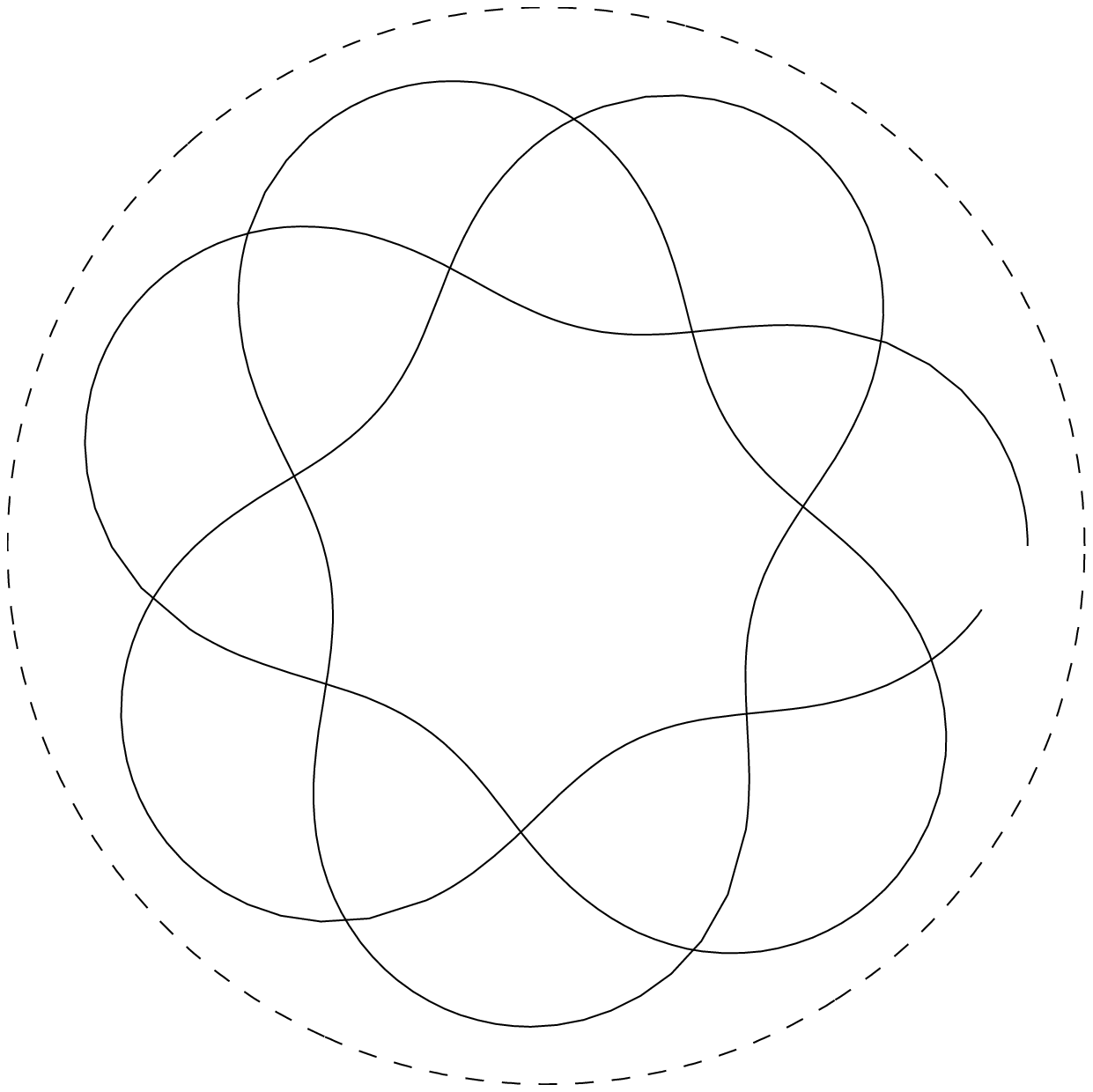}
\end{tabular}
\end{center}
\caption{Vortex trajectories under the Collie-Tong flow on $\Sigma_{2,0}$.
The curves show the trajectory of one of the vortices in each two
vortex motion, and illustrate a variety of behaviours: (a) scattering,
(b) an almost-circular bound orbit, (c) Larmor precession, (d) a cycloid-like
bound orbit.}
\label{fig3}
\end{figure}

Collie and Tong investigated the qualitative properties of this flow on
the moduli space of centred
Euclidean 2-vortices \cite{Collie:2008mx}, finding scattering
trajectories, bound orbits and bound orbits with Larmor precession. 
In this section we will make an analogous analysis of the flow on
$\Sigma_{2,0}$ numerically, and $\Sigma_{n,n-1}$ exactly. First, we note
that $\Sigma_{n,m}$ is a {\em totally geodesic} complex submanifold of
$\M_{n+m}$, so its Ricci form coincides with the restriction of the
Ricci form of $\M_{n+m}$ to $T\Sigma_{n,m}$. Hence, solutions of the
flow on $\Sigma_{n,m}$ are solutions of the flow on $\M_{n+m}$. Now on
any two-dimensional K\"ahler manifold, $\rho=\frac12\kappa\kf$ where
$\kappa$ is, as before, the scalar curvature, and $\kf$ is the K\"ahler form.
Hence (\ref{ctf}) (with $\lambda=2$) becomes
\beq\label{ctf2}
\nabla_{d/dt}\dot{\oa}-\kappa J\dot\oa=0.
\eeq
We have solved this equation numerically on $\Sigma_{2,0}$ for various initial
data using the ODE solver package in Matlab. The corresponding
vortex trajectories are depicted in figure \ref{fig3}. Appropriate
choices of initial data yield scattering trajectories, bound orbits and 
Larmor precession, as predicted by Collie and Tong in the
Euclidean context. One can also find complicated cycloid-like
trajectories.

\begin{figure}
\begin{center}
\begin{tabular}{cccc}
(a)&(b)&(c)&(d)\\
& & & \\
\includegraphics[scale=0.25]{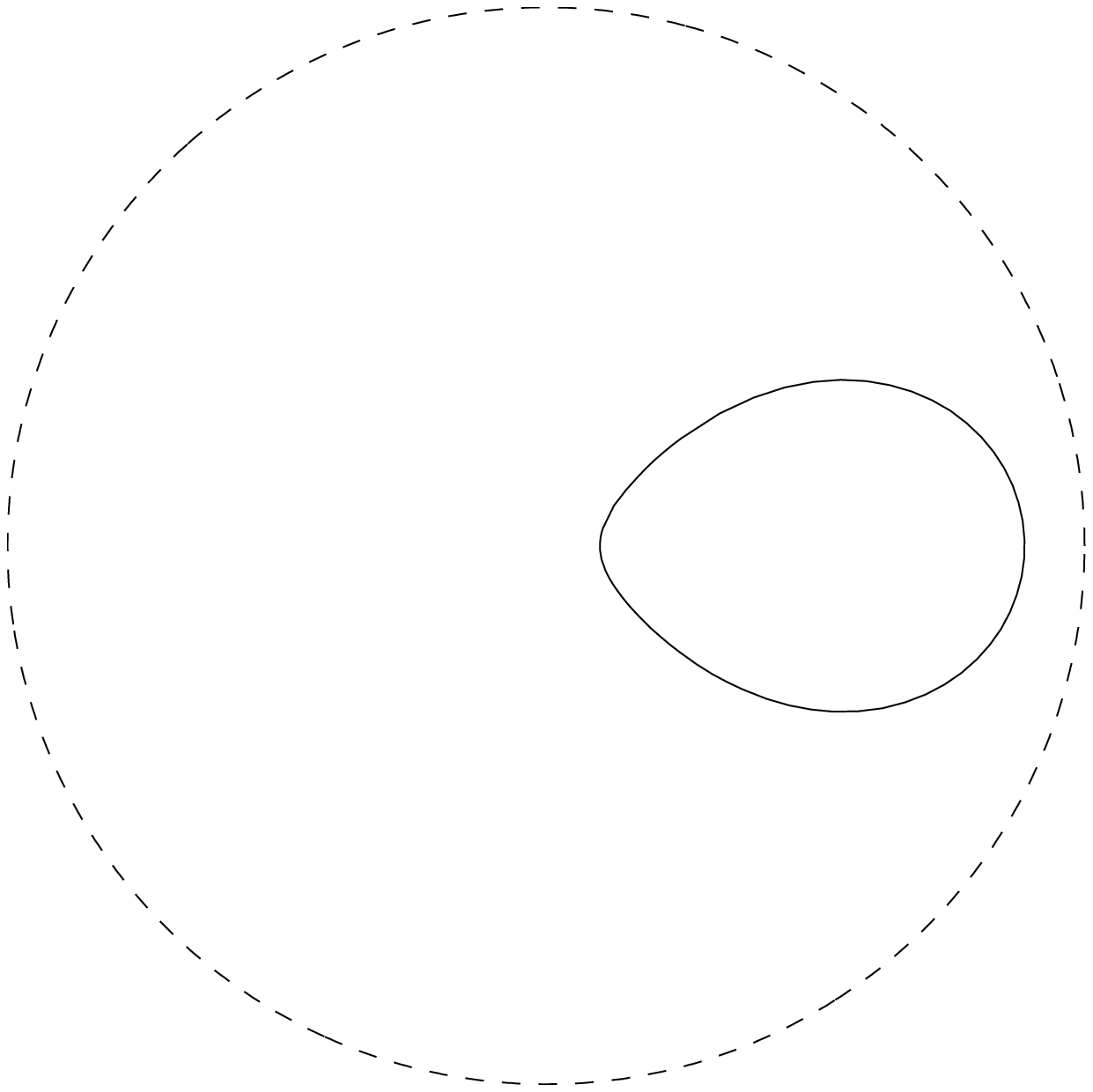}&
\includegraphics[scale=0.25]{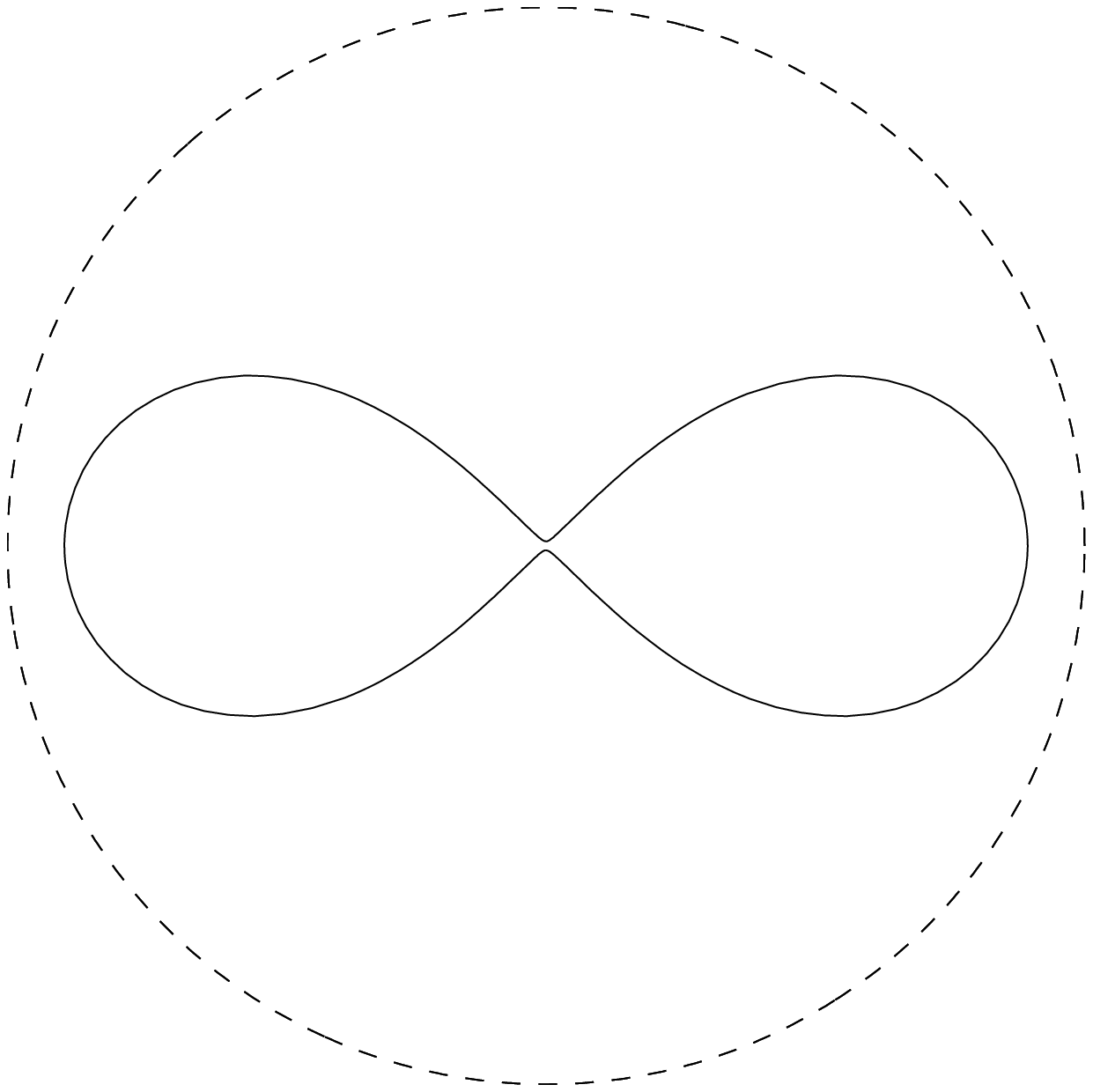}&
\includegraphics[scale=0.25]{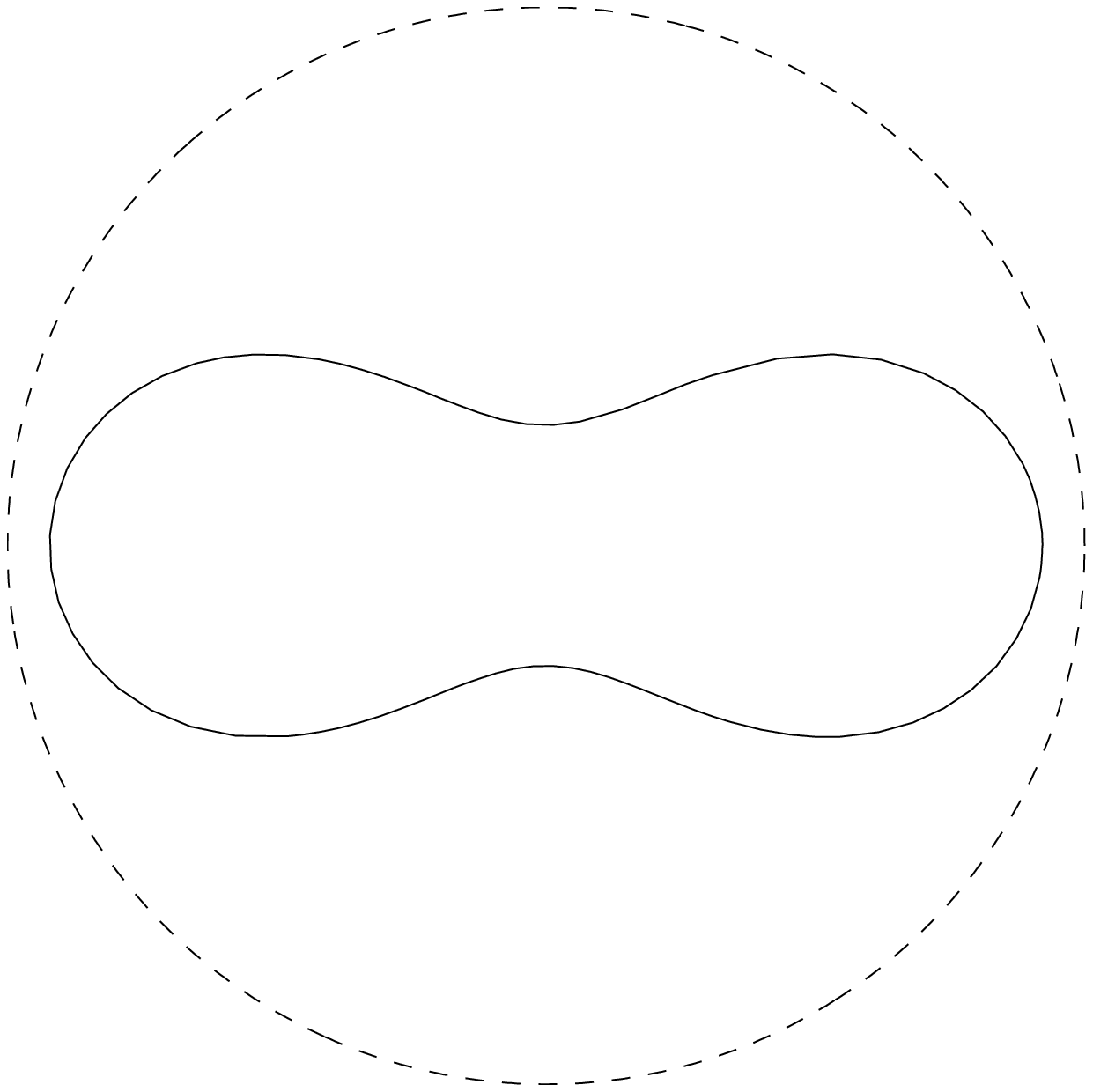}&
\includegraphics[scale=0.25]{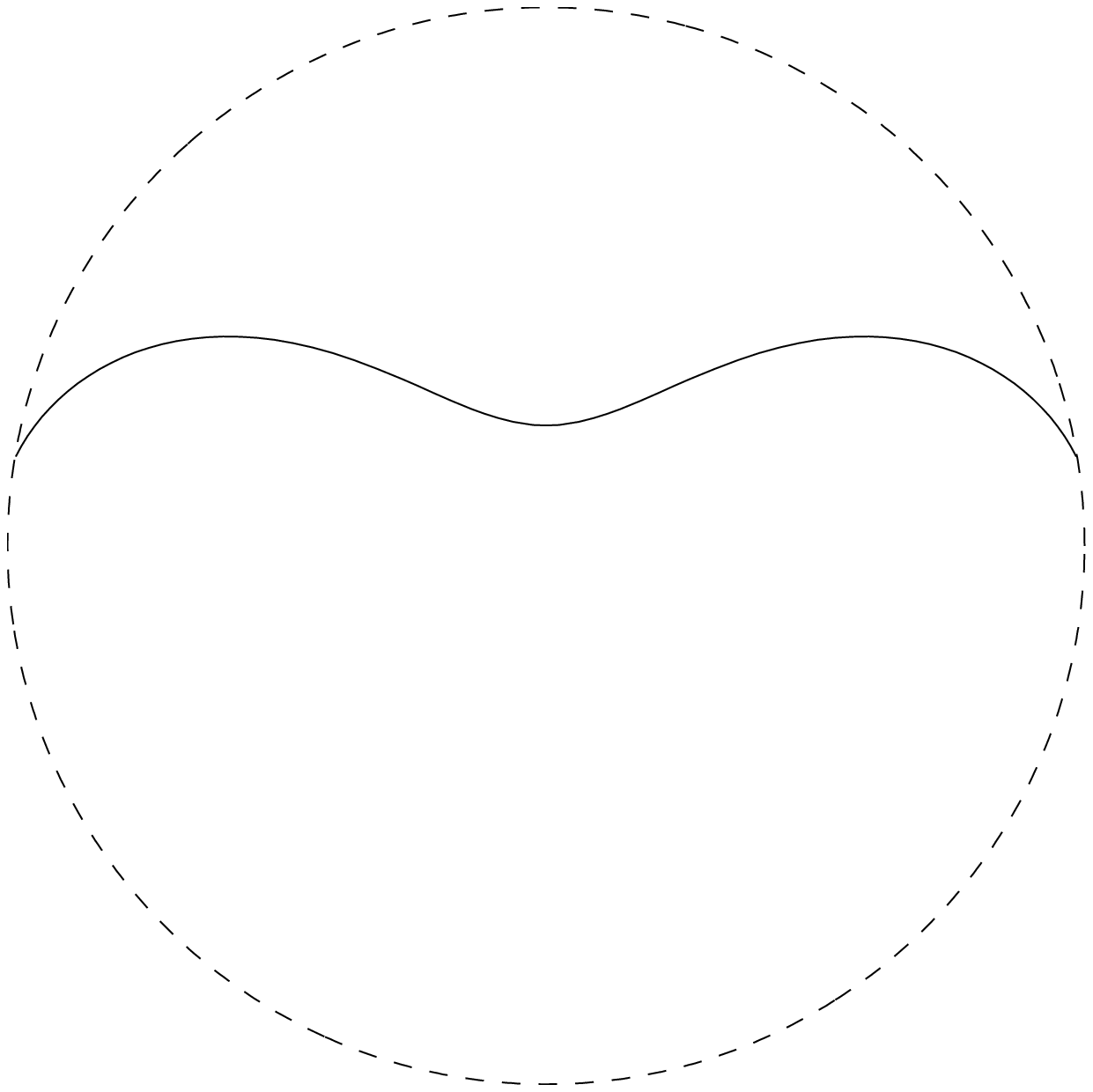}
\end{tabular}
\end{center}
\caption{Vortex trajectories under the Collie-Tong flow on $\Sigma_{2,1}$.
One vortex moves along the trajectory $\alpha(t)$ depicted, another
moves along $-\alpha(t)$ while a third vortex remains stationary at
the origin. In each case $\alpha(t)^2$ describes a circle with centre
in the unit disk. In case $(b)$ this circle passes through $0$ so the 
vortices undergo $90^\circ$ scattering at the origin. 
The circle captures $0$ in cases (c) and (d), but not in case (a).
Note that all
trajectories are periodic or escape to infinity.}
\label{fig4}
\end{figure}

We can make a more complete analysis of the flow on $\Sigma_{n,n-1}$,
since this is isometric to the hyperbolic plane of curvature $-1/3\pi n$,
so $\kappa$ is constant. By rescaling $t$ we can scale away $\kappa$ in
(\ref{ctf2}), so it suffices to consider the flow on, say, the hyperbolic
plane of curvature $-1$. In the upper half plane model this has metric
\beq
\gamma=\frac{1}{y^2}(dx^2+dy^2).
\eeq
To translate back to the disk model, we note that
\beq\label{mob}
\oa=\frac{z-i}{-iz+1},\qquad z=x+iy.
\eeq
The vortex trajectories are then given by the $n$-th roots of $\oa(t)$.
In the $z$ coordinate system (\ref{ctf2}) becomes
\bea
\ddot{x}-\frac{2}{y}\dot{x}\dot{y}+\dot{y}&=&0,\nonumber\\
\ddot{y}+\frac{1}{y}(\dot{x}^2-\dot{y}^2)-\dot{x}&=&0,
\eea
which has two conserved charges,
\beq
E=\frac{1}{2y^2}(\dot{x}^2+\dot{y}^2),\qquad
P=\frac{\dot{x}-y}{y^2}.
\eeq
Consider the trajectory with initial data $z(0)=i$,
$\dot{z}(0)=iv$ for $v>0$. This has $P=-1$, so its projection to the
$(y,\dot{y})$ phase plane is the energy level curve
\beq
\dot{y}^2=y^2(v^2-(1-y)^2).
\eeq
For $0<v<1$ the trajectories are periodic, with $1-v\leq y(t)\leq 1+v$  
while for $v\geq 1$ the trajectories are unbounded, escaping from and to
the boundary at infinity ($y=0$) as $t\ra\pm\infty$. Between critical points
of $y(t)$, we can determine the trajectory $x(y)$ by solving
\beq
\frac{dx}{dy}=\frac{\dot{x}}{\dot{y}}
=\pm\frac{1-y}{\sqrt{v^2-(1-y)^2}}.
\eeq
One finds that
\beq
(x+v)^2+(y-1)^2=v^2
\eeq
so the trajectory is a circle of radius $v$ centred on $(-v,1)$, or, if
$v\geq 1$, the portion of this circle in the upper half plane. 
Note, in the latter case, that the trajectory intersects the 
boundary acutely, not orthogonally, though the trajectories tend to a
geodesic as $v\ra\infty$, as one would expect.
By acting on this one-parameter family of circles with the isometry
group of the hyperbolic plane ($SL(2,\R)$ acting by fractional linear
transformations, in the upper half plane model), we obtain all circles
centred in the upper half plane (note that geodesics have centres on
the boundary $y=0$). Under the M\"obius transformation (\ref{mob}),
these are mapped to circles with centres in the unit disk, and every such
(arc of a) circle is a trajectory of the flow. The corresponding
vortex trajectories in $\X$ (in the disk model) are then obtained 
by taking $n$-th roots. 
Clearly the behaviour is much simpler than that found
numerically in $\Sigma_{2,0}$: all trajectories are either periodic or
unbounded (both as $t\ra\infty$ and $t\ra-\infty$).
Some examples are depicted in figure \ref{fig4}.

It would be interesting to make a systematic analysis of the flow
(\ref{ctf}) on a general K\"ahler manifold. One expects, for example,
that the flow is complete if and only if the manifold is metrically complete.
A natural generalization of the case $\Sigma_{n,n-1}$ considered in detail
above would be to assume that $\M$ is a homogeneous Einstein manifold,
for example $\M=\CP^N$.
This is potentially relevant to vortex dynamics on the two-sphere
close to the Bradlow limit \cite{Baptista:2002kb}.

\section*{Acknowledgements} 

SK would like to thank Nick Manton for helpful discussions at an early
state of the project.


\begin{thebibliography}{1}

\bibitem{Baptista:2002kb}
J.~M. Baptista and N.~S. Manton, {\em {The dynamics of vortices on S**2 near
  the Bradlow limit}\/}, J. Math. Phys. {\bf 44}: 3495--3508 ({\bf 2003}),
  {\ttfamily{<hep-th/0208001>}},

\bibitem{Chen:2004xu}
  H.~Y.~Chen and N.~S.~Manton, {\em {The Kaehler potential of Abelian
  Higgs vortices}\/}, 
  J.\ Math.\ Phys.\  {\bf 46} 052305 ({\bf 2005}),
  {\ttfamily{<hep-th/0407011>}},   

\bibitem{Collie:2008mx}
B.~Collie and D.~Tong, {\em {The Dynamics of Chern-Simons Vortices}\/}, Phys.
  Rev. {\bf D78}: 065013 ({\bf 2008}), {\ttfamily{<hep-th/0805.0602>}},

\bibitem{Kim:2002qma}
Y.~Kim and K.~Lee, {\em {First and second order vortex dynamics}\/}, Phys. Rev.
  {\bf D66}: 045016 ({\bf 2002}), {\ttfamily{<hep-th/0204111>}},

\bibitem{Krusch:2005wr}
S.~Krusch and P.~Sutcliffe, {\em {Schr{\"o}dinger-Chern-Simons Vortex
  Dynamics}\/}, Nonlinearity {\bf 19}: 1515--1534 ({\bf 2006}),
  {\ttfamily{<cond-mat/0511053>}},

\bibitem{McGlade:2005ug}
J.~A. McGlade and J.~M. Speight, {\em {Slow equivariant lump dynamics on the
  two-sphere}\/}, Nonlinearity {\bf 19}: 441 ({\bf 2006}),
  {\ttfamily{<hep-th/0503086>}},

\bibitem{Manton:2004tk}
N.~S. Manton and P.~Sutcliffe, {\em Topological solitons\/}, Cambridge
  University Press, Cambridge, U.K.
 ({\bf 2004}).

\bibitem{romthe} N.~M.~Romao, Ph.D. Thesis, University of Cambridge,
2002.

\bibitem{Romao:2004gc}
N.~M. Romao, {\em {Dynamics of CP(1) lumps on a cylinder}\/}, J. Geom. Phys.
  {\bf 54}: 42--76 ({\bf 2005}), {\ttfamily{<math-ph/0404008>}},

\bibitem{Samols:1991ne}
T.~M. Samols, {\em Vortex scattering\/}, Commun. Math. Phys. {\bf 145}:
  149--180 ({\bf 1992}),

\bibitem{Strachan:1992fb}
I.~A.~B. Strachan, {\em {Low velocity scattering of vortices in a modified
  Abelian Higgs model}\/}, J. Math. Phys. {\bf 33}: 102--110 ({\bf 1992}).


\end{thebibliography}
\end{document}